# Cryptanalysis of a Chaos-Based Fast Image Encryption Algorithm for Embedded Systems

Imad El Hanouti[1] · Hakim El Fadili[1] · khalid Zenkouar[2]

**Abstract**

Fairly recently, a new encryption scheme for embedded systems based on continuous third-order hyperbolic sine chaotic system was proposed by Z. Lin et al. [27]. The cryptosystem's main objective is to provide a faster algorithm with lowest computational time in order to be qualified for use in embedded systems especially on a *program of UAV (unmanned aerial vehicle)* [1]. In this paper, we scrutinize the design architecture of this recently proposed scheme against conventional attacks e.g., chosen plaintext attack, differential attack, known plaintext attack. We prove in this paper that, negatively, the studied system is vulnerable. For differential attack, only two chosen plain images are required to recover the full equivalent key. Moreover, only one $3 \times 400$ size image is sufficient to break the cryptosystem under chosen plaintext attack considering stability of sort algorithm. Therefore, the proposed scheme is not recommended for security purposes.

## 1. Introduction

Since its emergence in the 1960s [28], chaos theory has rapidly found so many applications in different fields [22-26]. Chaos-based cryptosystems [2-6] in which designers use chaotic systems as entropy source, has been widely studied in the literature since the early 1990s inspired by the strong similarity between chaos and cryptography [11]. However since then, so many cryptanalysis against chaos-based cryptography systems has been carried out [7-10]. Hence, many guidelines, requirements and recommendations have been proposed to ensure some basic security levels [11-12]. Even if, almost all proposed cryptosystems reassure the public community and reinforce their proposed systems by passing several randomness, statistical and empirical tests (e.g NIST statistical tests [29], histogram analysis, correlation analysis, UACI and NPCR [13]), those systems have been broken later by conventional cryptography attacks (e.g. chosen plaintext attack [14], known plaintext attack [15], differential attack [16] and so on).

[1] I. El Hanouti (✉) · H. El Fadili
Computer Science and Interdisciplinary Physics Laboratory (LIPI), National School of Applied Sciences, SMBA University, Fez, Morocco
Email : imad.elhanouti@usmba.ac.ma
[2] K. Zenkouar
Laboratory of Intelligent Systems and Application (LSIA), Faculty of Sciences and Technology, SMBA University, Fez, Morocco

The main motivations of many authors behind looking for new cryptosystems for multimedia data is that this kind of data generally tends to dominate the digital world [17-18] in addition to the fact that they have many specific characteristics. It can be summarized as follow: a strong redundancy between uncompressed neighboring elements (correlation), a bulk size of data, and some other special requirements like the necessity of fast encryption algorithms. All those specific characteristics, among others, limit the use of standard and conventional cryptosystems for this particular data type [19]. This paper could be seen as yet another counter example attempts to demonstrate the non-sufficiency of basic tests used by chaos-based cryptosystem designers to evaluate the security of their systems.

In this paper, a recently proposed cryptosystem for embedded systems is cryptanalyzed. The system under study is based on a continuous third-order hyperbolic sine chaotic system to build up a confusion/diffusion scheme with low computational complexity and fast execution time [27]. Although the authors literally announce that this "*proposed encryption scheme can resist known attacks, such as known-plaintext attacks, chosen ciphertext attacks, statistical attacks, differential attacks, and various brute-force attacks.*" and thus, "*Such encryption method with high quality and efficient could be appropriately applied in practice*" [27], we prove the opposite. Therefore, the encryption scheme is proven to be weak against differential, chosen plaintext and known plaintext attacks, and by the way it is not suitable for security and cryptographic use.

For the proposed differential attack, we will prove that only two chosen plain images are sufficient to recover the encryption equivalent key. Moreover, taking into consideration the stability of sort algorithm, an optimized chosen plaintext attack requires only one $3 \times 400$ size chosen image to break the system and recover the equivalent key for any image size. Additionally, statistical analysis of pairs of plain/cipher images in a known plaintext attack scenario could lead to the recovery of part or the entire equivalent key.

The rest of this paper is organized as follow; the second section presents the algorithm specifications and a concise description of all processes of encryption. The third section checks the system under differential attack. A chosen plaintext attack is performed in section 4. Section 5 discusses the applicability of known plain text attack on the studied scheme. Finally, section 6 concludes the paper.

## 2.    Description of the encryption process

The cryptosystem under study is designed to encrypt an $M \times N$ ($height \times width$) gray scale image according to a simple confusion/diffusion structure using a continuous third-order hyperbolic sine chaotic system [20] as base for chaotic key sequence generation as:

$$\dddot{x} + 0.75\ddot{x} + \dot{x} + 1.2 \times 10^{-6} \times \sinh\left(\frac{x}{0.026}\right) = 0 \qquad (1)$$

The sequence -directly generated- from this chaotic system is not random enough (as claimed by the main article authors [27]). Thus, in order to enhance the randomness of generated sequence the authors mixed up two elements of Eq(1) to generate a mixed sequence[3] $\{m(i)\}_{i=1}^{k}$, this operation is called *mixed sequence operation*:

$$\begin{cases} m(2n) = \dot{x}(2n) \\ m(2n+1) = \ddot{x}(2n+1) \end{cases}, n \in \mathbb{N} \quad (2)$$

Furthermore, to decorrelate the chaotic sequence and improve its randomness, the authors of the scheme has added one more enhancement to the sequence $\{m(i)\}_{i=1}^{k}$ to obtain a new pseudo-random sequence $\{c(i)\}_{i=1}^{k}$, this operation is called *decorrelation operation*:

$$c(i) = m(i) \times 10^6 - \lfloor(m(i) \times 10^6)\rfloor, i = 1 \sim k \quad (3)$$

Where $\lfloor . \rfloor$ denotes *floor* function, which takes the greater integer that is not greater than $(.)$.

The exponent 6 in $10^6$ is chosen such that the generated sequence $\{c(k)\}_{i=1}^{k}$ passed some NIST randomness test[4] [27_table3].

In order to reduce computational complexity and lighten the system to be more suitable for embedded systems, the authors fixed the length of the generated sequence (after some several statistical tests[5]) to 400 whatever the size of the image to encrypt is.

The encryption process could be described in three main steps:

1) Chaotic key sequence generation:
   Generate $\{c(i)\}_{i=1}^{k}$ the chaotic sequence with a fixed length of $k = 400$ elements as mentioned above, the secret key which we will refer to as *master secret key* is a triplet of three float numbers $(\dot{x}_0, \ddot{x}_0, \dddot{x}_0) \in [0,1]^3$.
2) Permutation process:
   This operation is divided into two main permutations: column transformation and row transformation. In column transformation each row is replaced with another one row, the index of the row is determined by a vector of indexes that determines the indexes of the sorted chaotic sequence $\{c(i)\}_{i=1}^{k}$ in ascending order ( as described in [27_appendix-A-]). In similar way, for row transformation each column is replaced with another column in respect to the vector of indexes of the sorted chaotic key sequence as done in column permutation stage.
   We can reformulate the permutation stage mathematically as follow:

$$Ip(i,j) = I(i^*, j^*) \quad (4)$$

Such that $I = \{I(i,j)\}_{1 \leq i,j \leq M.N}$ is the original image and $Ip = \{Ip(i,j)\}_{1 \leq i,j \leq M.N}$ is the permuted image, $i^* = p_c(i)$, $j^* = p_r(j)$, and:

$$\begin{cases} p_c = \{p_c(i)\}_{1 \leq i \leq M} \\ p_r = \{p_r(j)\}_{1 \leq j \leq N} \end{cases} \quad (5)$$

---

[3] Some notations from the original paper are modified without affecting its main meaning.
[4] NIST SP 800-22 test is performed
[5] Statistical tests include NPCR and UACI tests only!

where $\{c(i)\}_{i=1}^k$ is sorted in ascending order to get the permutation maps $\{p_c(i)\}_{1 \leq i \leq M}$ and $\{p_r(j)\}_{1 \leq j \leq N}$. Note that if the length of column or row is greater than 400, the sequence c is extended in a cyclic way such that for any integer $i, c(i) = c(mod(i, 400))$ where $mod(i, 400)$ denotes modulo reduction of $i$ by the modulus 400.

3) Substitution process:

Before starting the substitution process, some basic modifications must be carried out on the chaotic sequence. First of all, a new sequence called $\{c'(i)\}_{0 \leq i \leq 399}$ is calculated such that:

$$c'(i) = mod(c(i) \times 10^8, 256), \quad i = 0 \sim 399 \qquad (6)$$

Then a final sequence $\{s(i)\}_{0 \leq i \leq M.N-1}$ is obtained by extending the sequence $c'$ such that:

$$s(i) = c'(mod(i, 400)), \quad 0 \leq i \leq M.N - 1 \qquad (7)$$

This pseudorandom sequence is used to mask pixel values in order to obtain the final encrypted image $I'$:

$$I'(i,j) = Ip(i,j) \oplus s\left(mod((i-1)*N+j, 400)\right) \qquad (8)$$

Thus,

$$I'(i,j) = I(i^*, j^*) \oplus s\left(mod((i-1)*N+j, 400)\right) \quad i = 1 \sim M, j = 1 \sim N \qquad (9)$$

where $i^* = p_c(i), j^* = p_r(j)$ and $s(0) = s(400)$

$I'$ is the encrypted image, $s$ is defined in Eq(7) and $\oplus$ denotes *XOR* (bitwise exclusive OR, if applied on two matrix with the same size then it is calculated element wise).

*Remark:* in [27] the authors perform a 200 cyclic shift for one time to the masking sequence $s$ before the masking process, however, this operation do not affect the security of the system nor the cryptanalysis results, thus we ignore it in the rest of this paper.

## 3. Differential attack

In image cryptology, a simple scenario of differential attack is where the attacker has access to two plaintexts[6] denoted as $I_1$ and $I_2$ and their corresponding ciphertexts denoted as $I'_1$ and $I'_2$. The attack's objective is to construct a relationship between the difference of plaintexts $\Delta(I_1, I_2)$ and the corresponding difference in ciphertexts $\Delta(I'_1, I'_2)$.

The proposed attack is a typical *divide-and-conquer* attack which aims at first to recover the permutation map used for row and column permutation then extracting the pseudorandom sequence used for substitution. With only two chosen plaintexts and with the application of differential attack, we could recover the full map used for permutation due to the existence of some critical weaknesses in the "*confusion*" process design because of reusing the key including:

1) The same permutation map is used for both columns and rows, thus knowing the map of $Max(M, N)$ is equivalent to breaking both column permutation and row permutation.

***Proposition1:***

---
[6] In this paper the two terms "plaintext" and "plain image" are used equivalently, same thing for "ciphertext" and "cipher image"

Without loss of generality, let $Max(M,N) = N$ and $\{p_r(k)\}_{k=1}^{N}$ be the permutation map for row transformation, the permutation map for column transformation $\{p_c(k)\}_{k=1}^{M}$ could be extracted as: $p_c = \{p \in p_r \backslash p \leq M\}$ provided that the initial order of $\{p_r(k)\}_{k=1}^{N}$ is conserved.

2) If $Max(M,N) > 400$ then the pseudorandom sequence $\{c(i)\}_{i=0}^{399}$ used to generate permutation maps is extended in a cyclic way. Thus, for any integer $i \geq 400$: $c(i) = c(mod(i, 400))$, taking in consideration that the algorithm used to sort the sequence $c(k)$ could be *stable*[7], recovering the permutation map for only 400 elements is enough to deduce any map length.

***Proposition2****:*

Let $c' = \{c'(k)\}_{k=0}^{399}$ be the chaotic sequence as generated by Eq (5), for any integer $K > 400$ we have:

$c^* = \{c^*(k) = c'(mod(k, 400))\}_{k=0}^{K}$, and $p = \{p(i)\}_{i=0}^{399}$ is the permutation map generated from the chaotic sequence $c^*$, thus the permutation map $PK$ for K elements could be deduced by inserting every element $k \in \{400, .., K\}$ just after the element with value $mod(k, 400)$ in $p$:

$$PK = \{PK(i)\}_{i=0}^{K-1}$$

Where $PK(i) = \begin{cases} p(i), & \text{if } mod(i,2) = 0 \\ p(i-1) + 400, & \text{if } mod(i,2) = 1 \end{cases}, i = 0 \sim K-1$

3) The maximum number of required chosen plaintexts [21] to recover the permutation map is calculated by $\lceil \log_{256}(Max(M,N)) \rceil$, (where $\lceil . \rceil$ denotes *ceil* function, the least integer that is greater than or equal to (.)), considering p*roposition2*, one needs only a maximum of $\lceil \log_{256}(400) \rceil = \lceil 1.08 \rceil = 2$, however, we need only one chosen plain text to recover the permutation map as we will demonstrate in the remaining of the present section.

Thus, if we do cancel out XOR masking operation, the scheme will be equivalent to a permutation only encryption system, one can easily figure out from Eq (9) that if we XOR two ciphertexts:

$$\Delta(I'_1, I'_2) = I'_1 \oplus I'_2 \tag{10}$$

Then,

$$I'_1(i,j) \oplus I'_2(i,j) = I_1(i^*, j^*) \oplus I_2(i^*, j^*), i = 1 \sim M, j = 1 \sim N \tag{11}$$

Where $i^* = p_c(i)$ and $j^* = p_r(j)$, thus, only permutation operations remain. Once the permutation map is broken, the XOR substitution equivalent key could be extracted easily. In the following, the full attack is described step by step:

- Breaking the permutation map $p$ :
  As mentioned before, one need only to break whether column permutation or row permutation (the dimension of the maximum length) to deduce the full permutation map. Let us considering a square image with size $Max(M,N) \times Max(M,N) = N \times N$, thus $p_c = p_r = p$. If we choose the two plain images $I_1$ and $I_2$ in such a way to have an unambiguous pattern in each row of $I_1 \oplus I_2$, then the column permutation map could be recovered by looking for the permuted pattern in each row of the image $I'_1 \oplus I'_2$. Thus, we could recover $p$. For example, one could choose $I_1$ and $I_2$ such as:

$$I_1(i,j) \oplus I_2(i,j) = \begin{cases} 255, & i \leq j \\ 0, & i > j \end{cases}, i = 1 \sim N, j = 1 \sim M \tag{12}$$

---

[7] The sort algorithm is stable if it conserves the initial order of elements after resorting.

The number of white pixels (255 value pixel) in each row of $I'_1 \oplus I'_2$ defines simply the permutation key $p$ as:

$$p(j) = \frac{1}{255}\sum_{i=1}^{N} I'_1(i,j) \oplus I'_2(i,j), \quad j = 1 \sim N \qquad (13)$$

Hence, we have recovered the full permutation map using differential attack with only two chosen plaintexts.

- Breaking the masking pseudorandom sequence $s$ :
  After recovering the permutation map, the recovery of substitution key is much easier, considering for example the image $I_1$, the pseudorandom sequence for XOR operation could be recovered as follow:

$$c(j) = I_1(i^*, j^*) \oplus I'_1(i,j) \quad where\ i^* = p(i), j^* = p(j), j = 1 \sim 400\ and\ i = 1 \qquad (14)$$

The described attack works for any image size. In image sizes where for example $M > N$ one could take an $M \times M$ square image and recover the permutation key for column transformation and then uses *proposition2* to deduce permutation key for row transformation. Thus we could recover the full equivalent key of encryption with only two chosen plain texts, using a differential attack. Fig. 1 illustrates this attack to decrypt the image in Fig. 1-c encrypted with a secret unknown master key (0.41337 ,0.11121 ,0.24494), all images used to illustrate this attack are $512 \times 512$ size images. Note that in this paper all images used for illustrations are from SIPI[8] database, and all random keys are generated using octave[9] *rand* function.

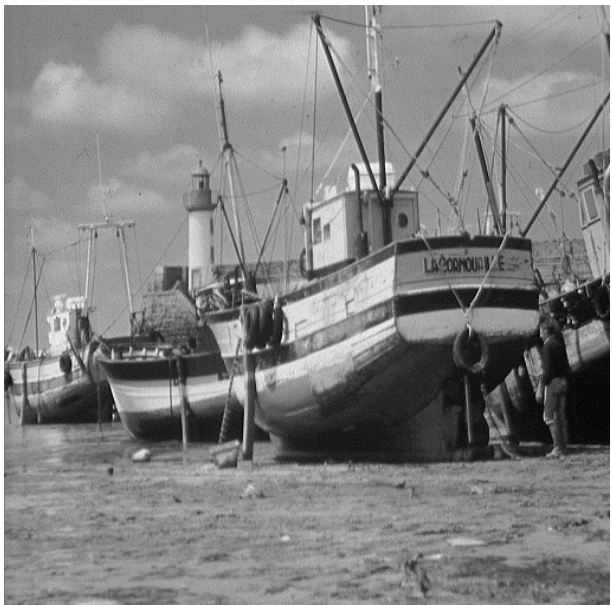 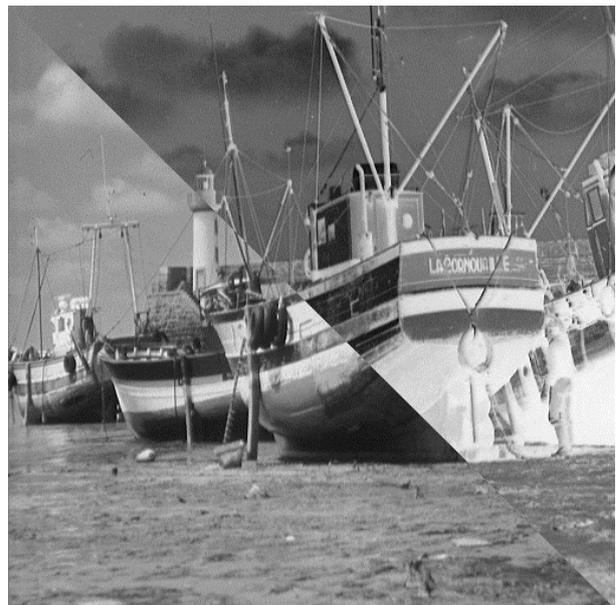

(a)  The first chosen plain image          (b) The second chosen plain image

---

[8] collection of digitized images at http://sipi.usc.edu/database/
[9] https://gnu.org/software/octave/

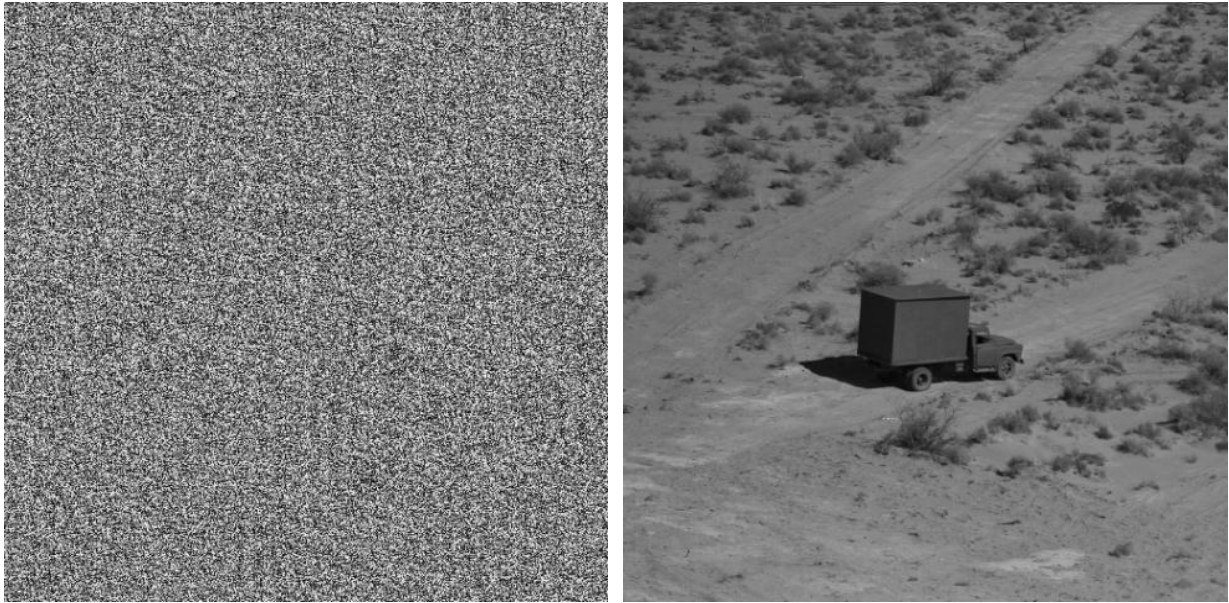

(c) The secret image            (d) The recovered image

**Fig. 1**: Illustration of the proposed differential attack

## 4. Chosen plaintext attack

We could optimize the above attack since we are dealing with a limited environment of resources and computational abilities, and only one required chosen plaintext with minimal size of $3 \times 400$ will be required to recover the full equivalent key of the encryption process considering the stability of sort algorithm.

Chosen plaintext attack is a well-known attack in which the attacker has the ability to choose some plaintexts to be encrypted with unknown key and get their corresponding ciphertexts. The objective of the attack is to determine the key, if a cryptosystem fails to resist the attack it is considered as unsecure.

In this attack we will take into consideration *propositions 1 and 2*. In order to attack the system we need only one known plaintext as we will demonstrate. This could be seen as a self-differential chosen plain text attack where we will use only one encryption operation to recover the equivalent key that could be used to decrypt any other cipher image encrypted by the same key. The proposed attack is operationally executed following the described steps:

- Breaking the substitution pseudorandom sequence $c$ :
  Since the permutation process is carried out before the substitution operation, if we could cancel out the permutation we could easily define the key sequence used to mask pixels by just XORing the ciphertext with the corresponding plaintext. If we choose the plain image $I$ such a way to have identical pixel value for at least one row with length of 400, then row permutation will be canceled for that row. Figuring out the corresponding row in the cipher image $I'$ will definitely allow us to recover the full pseudorandom sequence used in the substitution process $\{c(k)\}_{k=1}^{400}$.
- Breaking the permutation map $p$ :

Considering weaknesses outlined in the past section, we could in the same way as done before, recover the full permutation map. Thus, one could for example choose $I$ to be as:

$$I(i,j) = \begin{cases} I(1,:) = 0 \\ I(2,j) = mod(j, 256) + \left\lfloor \frac{j}{256} \right\rfloor, j = 1 \sim 400, i = 1 \sim 3 \\ I(3,j) = \left\lfloor \frac{j}{256} \right\rfloor \times 255 \end{cases} \quad (15)$$

which will define an image where the first row is all-black row and each column has a unique pattern.

**Recovering equivalent keys:**

Let $I'$ be the cipher image corresponding to $I$ described in (*). XORing $I'$ row-wise to identify the row $z$ which correspond to the black row in $I$, the substitution pseudorandom sequence could be recovered as:

$$c(i) = I'(z, i), \ i = 1 \sim 400 \quad (16)$$

Let $I^*$ be the substitution-only decrypted image:

$$I^*(i,j) = I'(i,j) \oplus c(j), \ i = 1 \sim 3, j = 1 \sim 400 \quad (17)$$

$I^*$ is a permuted image of $I$ according to the confusion operation of the encryption process, we could easily recover the row permutation:

$$p(j) = \sum_{i=1}^{3} I^*(i,j), \ j = 1 \sim 400 \quad (18)$$

Thus, only one known plaintext of minimum size of 3×400 is needed to recover the full equivalent key used for encryption of any image size. Note that the availability of this particular attack for any image size depends on the stability of the sort algorithm as mentioned in previous section. However, for any size smaller than $400 \times 400$ this particular attack could be implemented well and breaks the encryption scheme. Moreover, for an image size $q \times q$ smaller than $400 \times 400$ only one $3 \times q$ chosen image size is needed to break the key. Fig. 2 shows a demonstration of the attack to recover an encrypted $256 \times 256$ size image.

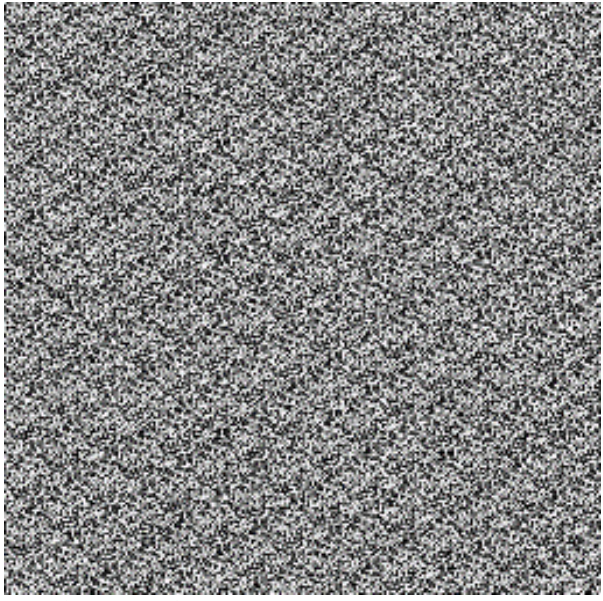 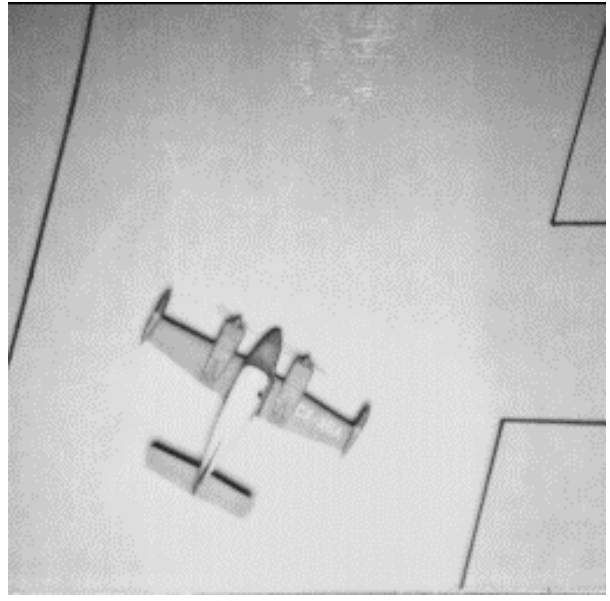

(a) The secret image　　　　　　　　　　　　(b) The recovered image
**Fig. 2**: Illustration of the proposed chosen plaintext attack using the chosen image described in Eq(15)

## 5.　Known plaintext attack

In this scenario, the attacker has intercepted some plaintexts with their equivalent ciphertexts, the objective is to recover the full or part of the key used to encrypt them. If one attacker could recover all or part of the key then the system is not secure.

For instance, considering one pair of plain/cipher text. Let $I$ be a plain image with size of $M \times 400.n$ where $n$ is an integer greater than zero, and $I'$ its corresponding cipher image. Therefore some very basic statistical tests could be carried out to reveal some parts of the equivalent key (or maybe all the key).

To break the permutation map (partially or even fully), one could for example count similar pixel values in each column (e.g. number of $k$ pixels in the $j^{th}$ column with the same pixel value: $\sigma(I(:,j) = k$ ), then if the count matches in the cipher image, the corresponding column is *elected* to be the permuted $j^{th}$ column in the image $I$. This way one could recover full or part of the permutation map $\{\hat{p}(k)\}_{k=1}^{400}$, and uses *proposition 2* (if applicable) to recover permutation map for any size.

After recovering parts or full permutation map, we could choose any row from $I'$ and recovering the substitution pseudo-random sequence as follow:

$$\hat{s}(j) = I(i^*, j^*) \oplus I'(i, j) \text{ with } i^* = \hat{p}(i), \text{ and } j^* = \hat{p}(j) \quad (19)$$

This is one way to do some statistical analysis in order to recover parts or full equivalent key used for encryption scheme from a known plaintext attack perspective.

## 6.　Conclusion

This paper evaluated the security of a lightweight image cryptosystem based on sine hyperbolic chaotic system that is tailored specifically for embedded systems. We have showed that, negatively, the scheme has several weaknesses against many attacks. For differential attack, only two chosen plaintexts are required to break the whole encryption scheme, moreover, we can optimize a chosen plaintext attack to break the encryption system with only one chosen image with optimal size of $3 \times 400$ considering the stability of sort algorithm. Thus, we could claim that: its main architectural design lies on vulnerable encryption scheme format and theoretically includes many algorithmic holes. Therefore, the system is cryptographically weak and it is not recommended at all for any security purpose.